\documentclass[journal, twoside]{IEEEtran}
\usepackage{indentfirst}
\usepackage{graphicx}
\usepackage{amsmath}
\usepackage{amssymb}
\usepackage{amsfonts}
\usepackage{mathrsfs}
\usepackage{leftidx}
\usepackage{color}
\usepackage{arydshln}
\usepackage{amsthm}
\usepackage{ragged2e}
\usepackage{cite}
\usepackage{enumerate}
\usepackage{longtable}
\usepackage{float}
\usepackage{stfloats}
\usepackage{hyperref}
\usepackage{algpseudocode}
\usepackage{algorithm}
\usepackage[caption=false,font=footnotesize]{subfig}
\usepackage{tabularx}
\usepackage{makecell}
\usepackage{url}
\usepackage{enumitem}
\usepackage{multirow} 
\usepackage{booktabs}
\theoremstyle{plain}

\usepackage{caption}

\newcolumntype{P}[1]{>{\raggedright\arraybackslash\footnotesize}m{#1}}
\newcolumntype{A}[1]{>{\centering\arraybackslash\footnotesize}m{#1}}

\usepackage[table,usenames,dvipsnames]{xcolor}
\definecolor{aa}{RGB}{175,238,238}
\definecolor{bb}{RGB}{255,255,255}

\usepackage{bm}
\usepackage{makecell}
\usepackage{siunitx} 
\usepackage{colortbl}
\usepackage{array}

\newcommand{\thickline}{\noalign{\hrule height 1.5pt}}
\newcommand{\thinline}{\noalign{\hrule height 0.5pt}}

\newcommand{\colwidth}{0.8cm}
\newcolumntype{C}[1]{>{\centering\arraybackslash}p{#1}}

\begin{document}

\title{IPRU: Input-Perturbation-based Radio Frequency Fingerprinting Unlearning for LAWNs}

\author{Ce Liu, Rui Meng,~\IEEEmembership{Member,~IEEE,} Yinqiu Liu,~\IEEEmembership{Member,~IEEE,}
Xiaodong Xu,~\IEEEmembership{Senior Member,~IEEE,} 

Yi Ma,~\IEEEmembership{Senior Member,~IEEE,} 
Rahim Tafazolli,~\IEEEmembership{Fellow,~IEEE,} 
and Ping Zhang,~\IEEEmembership{Fellow,~IEEE} 

\thanks{
\textit{(Corresponding author: Rui Meng and Xiaodong Xu.)}




Ce Liu, Rui Meng, Xiaodong Xu, and Ping Zhang are with the State Key Laboratory of Networking and Switching Technology, Beijing University of Posts and Telecommunications, Beijing 100876, China.
Yinqiu Liu is with the College of Computing and Data Science, Nanyang Technological University, Singapore.
Yi Ma and Rahim Tafazolli are with 5GIC \& 6GIC, Institute for Communication Systems (ICS), University of Surrey, Guildford, GU2 7XH, United Kingdom.

}}

\maketitle

\begin{abstract}
Radio Frequency Fingerprinting (RFF) is a key technology for identity authentication in wireless networks. However, due to the rapid dynamics of Autonomous Aerial Vehicles (AAVs) in low-altitude wireless networks, RFF models require parameter updates to maintain authentication performance, posing a major challenge to existing schemes. Conventional retraining approaches for handling departed or compromised AAVs are computationally prohibitive and risk retaining polluted features, which compromises both authentication security and user privacy. To address these limitations, we propose an Input-Perturbation-based RFF Unlearning (IPRU) scheme. By optimizing a universal Fingerprint Forget Vector (FFV) as a lightweight input perturbation, IPRU successfully erases the fingerprints of target AAVs without modifying the RFF model parameters, achieving an effective balance between efficient unlearning and preserved authentication performance. A combinatorial optimization strategy further enables multi-AAV forgetting on demand. The simulation results demonstrate that IPRU achieves 1.41\% unlearning accuracy, 99.41\% remaining accuracy, and 100\% resistance to membership inference attack, while running 5.79$\times$ faster than retraining and 2.1$\times$ faster than the baseline scheme.


\end{abstract}


\section{Introduction}


Radio Frequency Fingerprinting (RFF) has recently been considered a key technology for ensuring the access security of Low-Altitude Wireless Networks (LAWNs) \cite{zheng2025uav}. It constructs an unforgeable device ``identifier'' by extracting unique RF characteristics caused by hardware imperfections in transmitters, such as phase noise, power spectral distortion, and In-phase and Quadrature (IQ) imbalance \cite{mohammad2023learning-based}. Compared with traditional cryptography techniques, RFF avoids complex key distribution and maintenance, enabling lightweight authentication and endogenous security \cite{dhakal2025radio}. Therefore, it is particularly well-suited for Autonomous Aerial Vehicles (AAVs) with constrained energy and computational resources, effectively mitigating identity impersonation and access spoofing \cite{li2023physical}. For example, Teng \textit{et al.} \cite{teng2024exploit} presented a two-dimensional RFF scheme that utilizes carrier frequency offset and phase noise as AAV-specific hardware fingerprints. Compared with conventional single-feature or channel-dependent approaches, this scheme achieves improved robustness and authentication performance in highly dynamic LAWNs. Additionally, Zhou \textit{et al.} \cite{zhou2025lightweight} proposed a lightweight hybrid Convolutional Neural Network (CNN)-Transformer model for AAV-RFF, which effectively addressed the high computational complexity and limited adaptability of existing deep learning-based methods. It offers enhanced robustness and efficiency under low Signal-to-Noise Ratio (SNR) conditions, with limited training fingerprint samples, and in realistic multi-distance settings.

However, as the scale of LAWNs expands and the complexity of flight paths increases, existing RFF schemes struggle to adapt to highly dynamic environments where the topology changes frequently \cite{cheng2026apeg}. Taking AAV formation as an example, the frequent joining and leaving of individual AAVs in LAWNs is common. If the learned traces of the AAVs that have departed or compromised are not quickly forgotten, the model can retain the ``zombie'' or ``polluted'' features. This can lead to misidentification of legitimate users and enable attackers to persistently forge identities, severely compromising the accuracy of authentication \cite{liu2025threats}. In such contexts, existing techniques require retraining of the RFF model to accommodate dynamic AAV operations and ensure swarm security \cite{chen2025private}. In resource-constrained and latency-sensitive LAWNs, such high-overhead update mechanisms fail to meet ultra-low latency requirements, significantly limiting authentication efficiency.

To address the aforementioned challenges, we introduce machine unlearning techniques and propose an Input-Perturbation-based RFF Unlearning (IPRU) scheme. Machine unlearning aims to eliminate the influence of target data on model behavior and privacy leakage, rather than merely focusing on parameter updates \cite{sun2024forget}. Guided by this insight, IPRU achieves efficient fingerprint unlearning by generating universal unlearning perturbations without modifying model parameters, thereby reducing unlearning overhead while enhancing dynamic adaptability and privacy protection. The main contributions are summarized as follows.

\begin{itemize}
    \item We propose IPRU, a flexible access authentication scheme for LAWNs. By generating a Fingerprint Forget Vector (FFV) that serves as a universal input perturbation for RF fingerprints, the scheme enables efficient unlearning of targeted AAV fingerprints without modifying the RFF model parameters. It significantly decreases the success rate of Membership Inference Attack (MIA) post unlearning, while preserving the model’s identity authentication performance.
    \item We present an unlearning objective function that combines a retain loss with a regularization term to ensure efficient convergence of the unlearning process and minimal degradation of authentication performance. Additionally, we design an arithmetic combination mechanism for multi-AAV unlearning, which rapidly synthesizes new vectors from existing AAV-specific FFVs, enabling flexible unlearning for any AAV device.
    \item Extensive simulations on a public dataset \cite{shi2025rfuav} demonstrate that IPRU effectively unlearns target fingerprints and improves unlearning efficiency in both single- and multi-AAV unlearning scenarios, while maintaining high authentication accuracy for unrelated AAVs.
\end{itemize}
\section{System Model and Problem Formulation}

\subsection{System Model}


\begin{figure}
\centering
\includegraphics[width=0.45\textwidth]{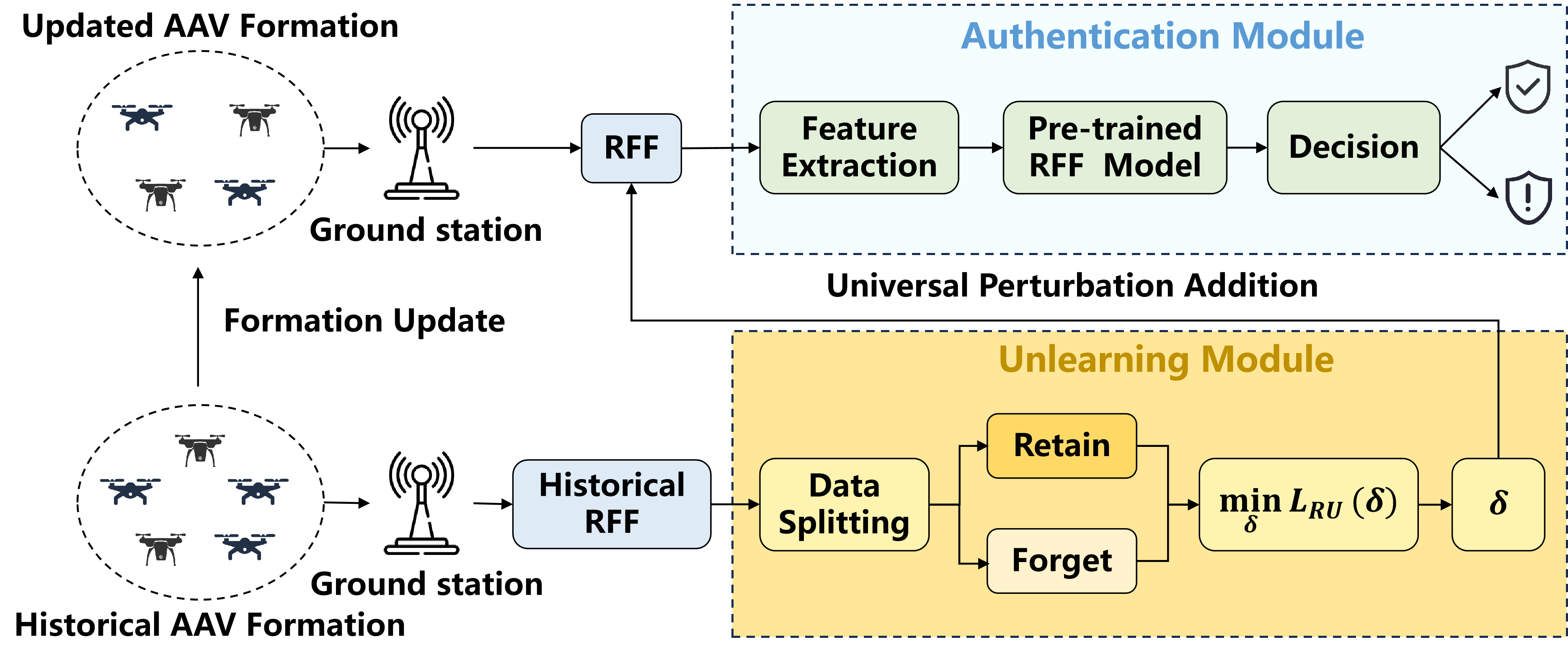} 
\caption{The overview of IPRU framework, where an authentication module verifies AAV identities via a pre-trained RFF model, a fingerprint unlearning module extracts specific FFV  $\boldsymbol{\delta}$ from partitioned historical data through targeted optimization, and the generated $\boldsymbol{\delta}$ erases the fingerprint traces of departed or compromised AAVs without altering model parameters.}
\label{fig1}
\end{figure}

Fig. \ref{fig1} illustrates the proposed IPRU framework. We consider an AAV formation characterized by frequent membership changes, in which individual AAVs may either leave the network normally or become compromised due to security vulnerabilities \cite{fugkeaw2026chaindrone}. At the ground station, RF signals are received and then passed to the subsequent authentication module for processing. This authentication module employs a pre-trained RFF model to verify identities based on hardware-specific fingerprints. To address the security risks posed by departed or compromised AAVs, the fingerprint unlearning module is triggered upon formation updates.

The fingerprint unlearning module operates as follows. It first partitions the historical RF fingerprint data into a forget set and a retain set. It then generates an AAV-specific forgetting perturbation $\boldsymbol{\delta}$, defined as the FFV, via targeted optimization. This perturbation serves as a universal input disturbance applied to RF signals at the receiver, effectively erasing its fingerprint traces from the RFF model without modifying the model parameters \cite{sun2024forget}. The resulting unchanged model continues to reliably authenticate legitimate AAVs while rendering signals from forgotten devices indistinguishable from those of unauthorized ones.



\subsection{Problem Formulation}
\label{definition} 

Let $\mathcal{F}_\theta$ denote the RFF model, where $\theta$ represents the model parameters. $\mathcal{F}_\theta$ is trained on a dataset consisting of data pairs of RF fingerprints and AAV identity labels, i.e., $D = \{\boldsymbol{x}_i, y_i\}_{i=1}^n$, where $\boldsymbol{x}_i$ represents the Short-Time Fourier Transform (STFT) sample generated from the original sampled signal ($I + jQ$) and contains RF fingerprint features; $y_i$ is the AAV identity label corresponding to $\boldsymbol{x}_i$, and $n$ denotes the total number of AAVs in the dataset. Following the standard machine unlearning setup \cite{gol2020eternal,fan2024salun}, we present a \textit{forget set} $D_f = \{\boldsymbol{x}_k, y_k\} \subseteq D$, where $k \in \{1, \dots, n\}$ denotes the index of the training sample to be forgotten. Correspondingly, the complement of $D_f$ is the \textit{retain set}, i.e., $D_r = D \setminus D_f$, whose information should remain unchanged in the model. Thus, the problem is to efficiently and effectively eliminate the influence of $D_f$ on the original RFF model $\mathcal{F}_\theta$ while maximizing the retention of authentication performance on $D_r$. Drawing on existing machine unlearning methods, this problem can be further formulated as 
\begin{equation}
\label{1}
\min_{\theta} \mathcal{L}_{RU}\left(\mathcal{F}_\theta; D_f, D_r\right),
\end{equation}
where $\mathcal{L}_{\text{RU}}$ denotes the unlearning loss function that depends on $D_f$ and $D_r$, and will be further formulated in Section \ref{method}.

\section{The Proposed IPRU Scheme}
\label{method} 

\subsection{Single-AAV Unlearning}
In the single-AAV unlearning scenario, for an STFT sample $\boldsymbol{x} \in D_f$ in the \textit{forget set}, the perturbed input is defined as $\boldsymbol{x}' = \boldsymbol{x} + \boldsymbol{\delta}$. IPRU performs unlearning through a universal perturbation $\boldsymbol{\delta}$ in the input space, rather than by directly updating the model parameters. Therefore, the optimization objective can be formulated as
\begin{equation}
\label{2}
\min_{\boldsymbol{\delta}} \mathcal{L}_{RU}\left(\boldsymbol{\delta}; \mathcal{F}_\theta, D_f, D_r\right).
\end{equation}
Following the forget vector optimization framework in \cite{sun2024forget}, for samples in the \textit{forget set} $D_f$, $\boldsymbol{\delta}$ is designed to cause the predictions of the perturbed samples to deviate from their original correct labels, so the forget-set loss function $\mathcal{L}_F(\cdot)$ can be derived as 
\begin{equation}
\label{3}
\begin{split}
\mathcal{L}_F\left(\boldsymbol{\delta}; \mathcal{F}_\theta, D_f\right) = \mathbb{E}_{(\boldsymbol{x},y) \in D_f} \max\left\{ f_{\mathcal{F}_\theta, y}(\boldsymbol{x}+\boldsymbol{\delta}) \right. \\
\left. - \max_{k \neq y} f_{\mathcal{F}_\theta, k}(\boldsymbol{x}+\boldsymbol{\delta}), -\tau \right\},
\end{split}
\end{equation}
where $(\boldsymbol{x},y) \in D_f$ denotes an RF signal sample pair in the \textit{forget set}, $\boldsymbol{x}$ is the STFT spectrogram generated from the original signal ($I + jQ$), and $y$ is the corresponding AAV label; $f_{\mathcal{F}_\theta, k}(\cdot)$ represents the logit output of the model $\mathcal{F}_\theta$ for class $k$. The term $\boldsymbol{x}+\boldsymbol{\delta}$ is the signal after applying the additive perturbation; $\tau \geq 0$ is a confidence margin parameter that controls the unlearning intensity. This loss function ensures that the optimization tends to terminate when the model’s prediction for the forgotten sample is converged to an incorrect device. 
The complete optimization objective is then given by
\begin{equation}
\label{4}
\begin{split}
\mathcal{L}_{RU}(\boldsymbol{\delta}; \mathcal{F}_\theta, D_f, D_r) = \alpha \mathcal{L}_R\left(\mathcal{F}_\theta, \{\boldsymbol{x}+\boldsymbol{\delta}, y\}_{(\boldsymbol{x},y) \in D_r}\right) \\
+ \beta \mathcal{L}_F(\boldsymbol{\delta}; \mathcal{F}_\theta, D_f) + \gamma \|\boldsymbol{\delta}\|_2^2,
\end{split}
\end{equation}
where $\alpha, \beta, \gamma > 0$ are balance hyper parameters; $\mathcal{L}_R(\cdot)$ denotes the cross-entropy loss, which is used to regularize the performance on the \textit{retain set} $D_r$; $\|\boldsymbol{\delta}\|_2^2$ is a regularization term that limits the perturbation amplitude to maintain the physical feasibility of the signal.

To validate the effectiveness of optimizing a universal $\boldsymbol{\delta}$ in RFF unlearning, we adopt a first-order analytical perspective \cite{goodfellow2015explaining}. For $(\boldsymbol{x}_i,y_i)$, we define the authentication loss as
\begin{equation}
\ell_i(\boldsymbol{\delta}) = \ell\!\left(\mathcal{F}_{\theta}(\boldsymbol{x}_i+\boldsymbol{\delta}),y_i\right), \quad \theta\ \text{is fixed},
\end{equation}
For a small update $\Delta\boldsymbol{\delta}$, the first-order Taylor expansion yields
\begin{equation}
\ell_i(\boldsymbol{\delta}+\Delta\boldsymbol{\delta}) \approx \ell_i(\boldsymbol{\delta}) + \nabla_{\boldsymbol{\delta}}\ell_i(\boldsymbol{\delta})^\top \Delta\boldsymbol{\delta},
\label{eq:taylor_delta}
\end{equation}
and the chain rule gives the gradient equivalence
\begin{equation}
\nabla_{\boldsymbol{\delta}}\ell_i(\boldsymbol{\delta}) = \nabla_{\boldsymbol{x}}\ell\!\left(\mathcal{F}_{\theta}(\boldsymbol{x}_i+\boldsymbol{\delta}),y_i\right)
\label{eq:grad_equiv}
\end{equation}
Thus, updating $\boldsymbol{\delta}$ is equivalent to input-gradient optimization. Under the stochastic gradient descent (SGD) \cite{amari1993backpropagation} update $\boldsymbol{\delta}\leftarrow \boldsymbol{\delta}-\eta\nabla_{\boldsymbol{\delta}}\mathcal{L}_{RU}$, the first-order approximation ensures that the objective decreases with a sufficiently small learning rate $\eta$, achieving the unlearn-retain trade-off while the perturbation energy is constrained by the $L_2$ regularization.

\begin{algorithm}[t!]
\caption{Unlearning Steps of the proposed IPRU scheme}
\label{algorithm1}
\textbf{Input:} Historical STFT spectrograms $\boldsymbol{x} \in \{D_f, D_r\}$; $\mathcal{F}_\theta$; $T_h$.

\textbf{Process:}
\begin{algorithmic}[1]
\State Freeze the original model weights $\theta$ of  $\mathcal{F}_\theta$;
\State Initialize the $\boldsymbol{\delta}$ as a trainable tensor with the same shape as $\boldsymbol{x}$;
\Repeat 
    \State Calculate objective loss by (\ref{4});
    \State Update the perturbation $\boldsymbol{\delta}$ via SGD;
    \State Apply $\boldsymbol{\delta}$ to the test set of STFT spectrograms and evaluate feedback metrics;
\Until{The metrics satisfy the specified thresholds $T_h$}
\If{Multi-AAV unlearning is required}
    \State Collect the precomputed class-level FFVs $\{\boldsymbol{\delta}_k\}_{k=1}^n$;
    \State Construct combined perturbation (\ref{eq8});
    \State Optimize coefficients $\boldsymbol{c}$ by (\ref{eq9});
\EndIf
\Statex
\textbf{Output:} The optimized perturbation $\boldsymbol{\delta}^*$ (Single-AAV) or $\boldsymbol{\delta}(\boldsymbol{c})^*$ (Multi-AAV)
\end{algorithmic}
\end{algorithm}

\subsection{Multi-AAV Unlearning}
Multi-AAV unlearning is achieved via the joint optimization of $\boldsymbol{\delta}$ based on the single-AAV unlearning strategy. We first precompute the class-level FFVs $\boldsymbol{\delta}_k$ for each AAV with index $k=1, \dots, n$: taking all samples of the target AAV as $D_f$ and the remaining samples as $D_r$, and independently solving $\mathcal{L}_{RU}(\cdot)$ to obtain the set $\{\boldsymbol{\delta}_k\}_{k=1}^n$. 
Under the assumption that the forgetting effects of different FFVs can be approximately linearly combined, without explicitly modeling potential nonlinear interactions among FFVs, a combined perturbation for multi-AAV unlearning is constructed as
\begin{equation}
\label{eq8}
\boldsymbol{\delta}(\boldsymbol{c}) = \sum_{k=1}^n c_k \boldsymbol{\delta}_k,
\end{equation}
where $\boldsymbol{c}=[c_1, \dots, c_n]^\top$ denotes the weight coefficients of each FFV, and the corresponding optimization objective is formulated as
\begin{equation}
\label{eq9}
\min_{\boldsymbol{c}} \mathcal{L}_{RU}\big(\boldsymbol{\delta}(\boldsymbol{c}); \mathcal{F}_\theta, D_f, D_r\big).
\end{equation}

Algorithm \ref{algorithm1} details the unlearning procedure of IPRU. The universal perturbation $\boldsymbol{\delta}$ is adaptively generated from the feature distributions of the forget and retain sets and precisely erases target RF fingerprint features via iterative optimization. $T_h$ denotes the performance threshold for unlearning.

\textit{Complexity Analysis:} Given an input STFT spectrogram of size $H \times H$, the backbone network is frozen in both single- and multi-AAV unlearning. Their overall time complexity over $E$ epochs and $N$ samples is $\mathcal{O}\!\left(E \cdot \frac{N}{B} \cdot \left(F(B) + G(B)\right)\right)$, where $F(B)$ and $G(B)$ denote the forward propagation and backward propagation costs with respect to the optimized variables for a mini-batch of size $B$, respectively. The corresponding additional trainable memory overhead is $\mathcal{O}(H^2)$ for single-AAV unlearning and $\mathcal{O}(n)$ for multi-AAV unlearning, where $n$ is the number of class-level FFVs.

\section{Simulation Results and Analysis}
\subsection{Simulation Parameters}
\subsubsection{Dataset}
We utilize the RFUAV dataset \cite{shi2025rfuav} to verify the performance of the proposed IPRU scheme. This dataset comprises approximately 1.3 TB of raw RF data collected from 37 distinct AAVs in real-world environments via Universal Software Radio Peripheral (USRP) devices, with the data stored following the standard IQ sampling format. It includes multiple types of Frequency-Hopping Spread Spectrum (FHSS) signals generated by AAVs during pairing and flight operations. 
We select 10 types of AAVs for experiments.

\subsubsection{Data Preprocessing}
For the 10 selected types of AAVs, the IQ components are combined into the form $(I + jQ)$ to reconstruct the complete sampled signals. Subsequently, STFT is adopted to convert the complex signals into 2D time-frequency spectrograms, which effectively highlight the time-frequency energy distribution characteristics of the signals and extract their inherent hardware fingerprints. The obtained spectrograms are uniformly cropped to $224 \times 224$ after normalization of amplitude to meet the input requirements of deep learning models. ResNet-18 \cite{he2016deep} is adopted as the RFF authentication network and trained on the preprocessed spectrogram dataset. The original trained authentication model attains a  classification accuracy of 99.703\%  on the test set.

\subsubsection{Validation Framework}
We conduct the validation of IPRU framework under two primary scenarios: single-AAV unlearning and multi-AAV unlearning. For single-AAV unlearning experiments, Retrain \cite{thudi2022unrolling} and SCRUB \cite{kurmanji2023towards} are used as baseline methods for comparative analysis with the proposed IPRU scheme. In the multi-AAV unlearning scenario, we further investigate the performance discrepancies between the single-vector and combinatorial unlearning strategies.

\subsubsection{Parameter Settings}
In the single-AAV unlearning experiments, all 10 AAVs are evaluated. For multi-AAV unlearning, we target the simultaneous unlearning of FLYSKY FS I6X and FUTABA-T14SG. To strike a balance between unlearning effectiveness and retention performance, the hyper parameters are set as detailed in Table \ref{table1}. The performance thresholds $T_h^\text{UA}$ and $T_h^\text{RA}$ are set respectively to 5\% and 98\%, while the learning rate $\eta$ is set at 0.01. Additionally, the batch size is configured to 256, and the maximum number of optimization iterations is capped at 40. All experiments are implemented on an NVIDIA RTX 6000 Ada Generation GPU with PyTorch 2.5.1 and CUDA 12.1.

\begin{table}[htbp]
  \centering
  \caption{Hyper parameter Settings}
  \setlength{\tabcolsep}{4pt}    
  \renewcommand{\arraystretch}{1.2} 
  \begin{tabular}{>{\centering\arraybackslash}p{4cm}|*3{>{\centering\arraybackslash}m{\colwidth}}}
    \thickline 
    \multicolumn{4}{c}{\cellcolor{gray!20}Single-AAV unlearning} \\ 
    \thinline 
    Device & $\alpha$ & $\beta$ & $\gamma$ \\  
    \thinline
    DAUTEL EVO nano   & 1 & 16 & 2 \\  
    DEVENTION DEVO    & 1 & 6  & 3 \\
    DJI AVATA2        & 1 & 6  & 3 \\
    DJI FPV COMBO     & 1 & 6  & 3 \\
    DJI MAVIC3 PRO    & 1 & 6  & 3 \\
    DJI MINI3         & 1 & 3  & 3 \\
    DJI MINI4 PRO     & 1 & 12 & 3 \\
    FLYSKY EL 18      & 1 & 12 & 3 \\
    FLYSKY FS 16X     & 1 & 6  & 3 \\
    FUTABA-T14SG      & 1 & 12 & 3 \\
    \thinline
    \multicolumn{4}{c}{\cellcolor{gray!20}Multi-AAV unlearning} \\
    \thinline
    Method & $\alpha$ & $\beta$ & $\gamma$ \\  
    \thinline
    Single-V & 1 & 6  & 3 \\
    Com-V    & 1 & 8  & 1 \\
    \thickline 
  \end{tabular}
  \label{table1}
\end{table}

\subsection{Performance Metrics}
Motivated by \cite{jia2023model}, we adopt the following metrics to evaluate the proposed IPRU scheme.

\subsubsection{Unlearning Accuracy (UA)}
UA quantifies the classification accuracy of the unlearned model on the test set for the forgotten classes. A lower UA value represents a more thorough unlearning effect, reflecting the generalized erasure capability of the perturbation vector for the RF fingerprints of target devices.

\subsubsection{Remaining Accuracy (RA)}
RA measures the classification accuracy of the unlearned model on the test set for the retained classes. A higher RA value indicates better retention of the authentication performance for non-target devices and demonstrates that the unlearning operation does not significantly compromise the model's generalization capability.

\subsubsection{MIA-Efficacy}
MIA-Efficacy is adopted to assess the model's privacy protection performance. First, we calculate the entropy of the model's output probability distributions for samples in the retain set and test set, then train a Support Vector Classifier (SVC) based on entropy differences to differentiate between member and non-member samples. The trained classifier is utilized to predict samples in the unlearned set: a proportion of samples predicted as non-members close to 1 indicates adequate unlearning, while a value near 0.5 implies incomplete unlearning and persistent leakage of the original training data.

\subsubsection{Run-Time Efficiency (RTE)}
RTE records the computation time required to achieve unlearning, measured in minutes.
   
\subsection{Simulation Results}

\subsubsection{Single-AAV}
Table \ref{table2} presents the average performance metrics for single-AAV unlearning. In terms of unlearning efficacy, compared with SCRUB, our IPRU scheme yields performance closer to that of Retrain, narrowing the performance gap with Retrain to a mere 0.64\%. The 100\% MIA-Efficacy further demonstrates the effectiveness of the unlearning process. Regarding RA, the IPRU scheme also meets the performance preservation requirements for the retain set, with its RA trailing that of SCRUB by a mere 0.6\%. To explore why IPRU exhibits a marginally lower RA than SCRUB, we analyze the authentication results in Table \ref{table3} using the FUTABA-T14SG unlearning case. The results indicate that the input perturbation slightly interferes with common foundational RF fingerprint features. Consequently, DJI AVATA2, DJI MINI4 PRO, and FLYSKY EL 18 experience minor accuracy degradations of 0.64\%, 0.91\%, and 0.69\%, respectively, relative to the SCRUB baseline. For operational efficiency, the IPRU scheme shows clear superiority, providing a runtime acceleration of approximately 5.79$\times$ over Retrain and 2.1$\times$ over SCRUB, respectively. Notably, owing to the highly concentrated intra-class distribution of RF fingerprints and the weak distinguishability between member and non-member samples, both SCRUB and the IPRU scheme achieve 100\% MIA-Efficacy, indicating a high level of privacy protection.

\begin{table}[t!]
  \centering
  \caption{Average performance metrics for single-AAV unlearning}
  \label{table2}
  \setlength{\tabcolsep}{0pt} 
  \renewcommand{\arraystretch}{1.2} 
  \begin{tabular}{>{\centering\arraybackslash}p{3cm}|*{4}{>{\centering\arraybackslash}p{1.3cm}}}
    \thickline 
    Metric & Original & Retrain & SCRUB & Ours \\ 
    \thinline 
    UA (\%)           & 99.68 & 0.00    & \textbf{2.38}    & \textbf{0.64}\\ 
    RA (\%)           & 99.70 & 99.79   & 99.79   & 99.19  \\
    MIA-Efficacy (\%) & 4.10  & 100.00  & 100.00  & 100.00 \\
    RTE (min)         & ---   & 110.45  & 40.06   & \textbf{19.08} \\
    \thickline 
  \end{tabular}
\end{table}

\begin{table}[t!]
  \centering
  \caption{Accuracy performance on different AAVs when unlearning FUTABA-T14SG}
  \label{table3}
  \setlength{\tabcolsep}{0pt}
  \renewcommand{\arraystretch}{1.2}
  \begin{tabular}{>{\centering\arraybackslash}p{3cm}|*{4}{>{\centering\arraybackslash}p{1.3cm}}}
    \thickline
    Device & Original & Retrain & SCRUB & Ours \\
    \thinline
    DAUTEL EVO nano   & 98.78  & 99.59  & 99.59  & 99.59  \\
    DEVENTION DEVO    & 99.10  & 99.10  & 99.10  & 99.10  \\
    DJI AVATA2        & 99.58  & 99.79  & 99.58  & \textbf{98.94}  \\
    DJI FPV COMBO     & 100.00 & 100.00 & 100.00 & 100.00 \\
    DJI MAVIC3 PRO    & 99.49  & 99.87  & 99.62  & 99.62  \\
    DJI MINI3         & 99.91  & 99.91  & 99.91  & 99.91  \\
    DJI MINI4 PRO     & 99.63  & 99.27  & 99.63  & \textbf{98.72}  \\
    FLYSKY EL 18      & 99.66  & 100.00 & 100.00 & \textbf{99.31}  \\
    FLYSKY FS 16X     & 99.64  & 99.64  & 99.64  & 99.64  \\
    FUTABA-T14SG      & 100.00 & 0.00   & \textbf{0.21}   & \textbf{0.21}   \\
    \thickline
  \end{tabular}
\end{table}

\begin{table}[t!]
  \centering
  \caption{Performance metrics for multi-AAV unlearning.
Single-V denotes multi-AAV unlearning realized by a single FFV, while Com-V represents the combinatorial strategy.}
  \label{table4}
  \setlength{\tabcolsep}{0pt}
  \renewcommand{\arraystretch}{1.2}
  \begin{tabular}{>{\centering\arraybackslash}p{3cm}|*{4}{>{\centering\arraybackslash}p{1.3cm}}}
    \thickline
    Metrics & Original & Retrain & Single-V & Com-V \\
    \thinline
    UA (\%)           & 99.87 & 0.00    & \textbf{0.80}  & \textbf{3.34}   \\
    RA (\%)           & 99.68 & 99.89   & 99.53   & 99.25  \\
    MIA-Efficacy (\%) & 1.60  & 100.00  & 100.00  & 100.00 \\
    RTE (min)           & --- & 143.29  & 36.68   & \textbf{21.43} \\
    \thickline
  \end{tabular}
\end{table}

\subsubsection{Multi-AAV} 
Next, after obtaining the pre-trained FFVs for each device, we apply the combinatorial unlearning method formulated in (\ref{eq8}) and examine its performance discrepancies relative to Retrain and the single-vector method, as shown in Table \ref{table4}. In multi-AAV unlearning, Com-V achieves comparable performance to Single-V. Specifically, with a UA of 3.34\%, Com-V undergoes merely a 0.28\% decrease in RA relative to Single-V. This marginal decrease in RA is primarily attributable to the fact that the combinatorial method is effectively equivalent to simultaneously imposing multiple perturbation directions within the input space, thereby more readily disrupting the shared features of the retained classes. Moreover, since Com-V only needs to optimize a small number of coefficients while reusing the pre-computed single-AAV FFVs, the optimization space is reduced. Consequently, operational efficiency is improved, achieving a speedup of 1.71$\times$ over the Single-V approach.


\begin{figure}[t!]
    \centering
    \setlength{\tabcolsep}{0.5pt} 
    
    \begin{tabular}{c @{\hspace{3pt}} ccccc}
        & \textbf{\fontsize{6pt}{6.6pt}\selectfont RFF Image} 
        & \textbf{\fontsize{6pt}{6.6pt}\selectfont Original Model} 
        & \textbf{\fontsize{6pt}{6.6pt}\selectfont Retrain} 
        & \textbf{\fontsize{6pt}{6.6pt}\selectfont SCRUB} 
        & \textbf{\fontsize{6pt}{6.6pt}\selectfont IPRU} \\
        
        \raisebox{-0.5\height}{\rotatebox{90}{\textbf{\tiny DJI FPV COMBO}}} &
        \raisebox{-0.5\height}{\includegraphics[width=0.17\linewidth]{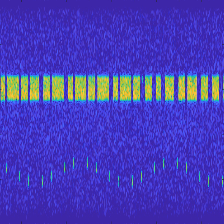}} &
        \raisebox{-0.5\height}{\includegraphics[width=0.17\linewidth]{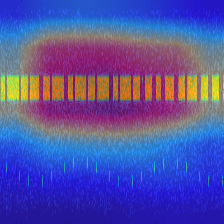}} &
        \raisebox{-0.5\height}{\includegraphics[width=0.17\linewidth]{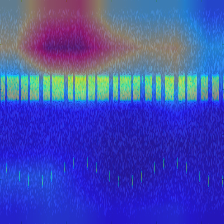}} &
        \raisebox{-0.5\height}{\includegraphics[width=0.17\linewidth]{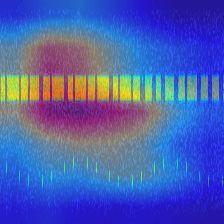}} &
        \raisebox{-0.5\height}{\includegraphics[width=0.17\linewidth]{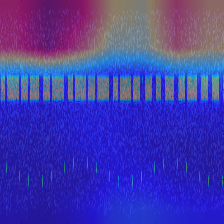}} \\
        
        \noalign{\vspace{1pt}} 
        
        \raisebox{-0.5\height}{\rotatebox{90}{\textbf{\tiny FLYSKY FS I6X}}} &
        \raisebox{-0.5\height}{\includegraphics[width=0.17\linewidth]{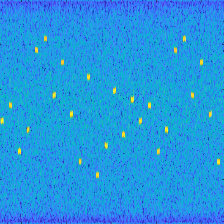}} &
        \raisebox{-0.5\height}{\includegraphics[width=0.17\linewidth]{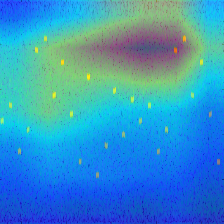}} &
        \raisebox{-0.5\height}{\includegraphics[width=0.17\linewidth]{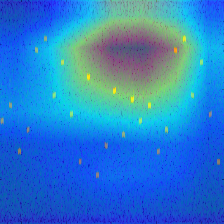}} &
        \raisebox{-0.5\height}{\includegraphics[width=0.17\linewidth]{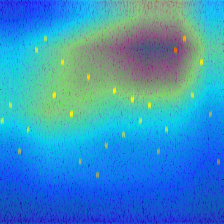}} &
        \raisebox{-0.5\height}{\includegraphics[width=0.17\linewidth]{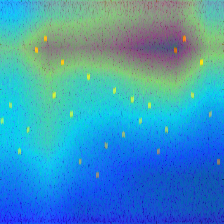}} \\
    \end{tabular}
    
    \caption{Grad-CAM visualization of model performance on the unlearning device (DJI FPV COMBO) and retaining device (FLYSKY FS I6X) under different unlearning methods. Red-highlighted regions indicate the areas that contribute most to the model's prediction.}
    \label{fig2}
\end{figure}

\subsubsection{Grad-CAM Studies}
As illustrated in Fig. \ref{fig2}, we employ gradient-weighted class activation mapping (Grad-CAM) \cite{selvaraju2017grad-cam} to examine the impact of the proposed IPRU and other baseline methods on model unlearning and utility retention. The Grad-CAM visualizations highlight the critical regions that dominate the model's prediction. The first row of Fig. \ref{fig2} shows the visualization results for the target unlearning AAV device (DJI FPV COMBO). Relative to the original model, the salient regions of both the Retrain and IPRU show clear and substantial shifts, which verifies their effectiveness in erasing the target device's RF fingerprints. In comparison, SCRUB generates a small shift in saliency maps, implying that the model fails to completely forget the corresponding target data. The second row of Fig. \ref{fig2} depicts the results for the retaining device (FLYSKY FS I6X). The salient regions of all unlearning methods are generally consistent with those of the original model, which confirms that the proposed method preserves robust authentication performance for non-target devices while completing the target device unlearning.

\section{Conclusions}
In this paper, we have applied input-perturbation-based machine unlearning to RFF in LAWNs and proposed a lightweight IPRU framework to meet the dynamic identity verification needs of AAV formations. By constructing a universal FFV, our method realizes efficient erasure of RF fingerprints for designated devices without modifying any model parameters, and strikes an excellent balance between unlearning efficacy and retention performance. The $\boldsymbol{\delta}$ optimization strategy ensures fast unlearning and stable system utility, and the combinatorial generation mechanism for multi-AAV further enhances the flexibility and operational efficiency of IPRU. Experimental results have demonstrated that IPRU achieves superior unlearning performance, high retention accuracy, and low computational overhead in both single-AAV and multi-AAV unlearning scenarios, verifying its effectiveness and practical value in low-altitude environments. In future work, we will enhance the robustness of the model against noisy interference and incorporate online update mechanism to enable continuous and fine-grained identity recognition.

\section{Acknowledgment}
We would like to sincerely thank Dusit Niyato from NTU for his contributions to this article.

\bibliographystyle{IEEEtran} 
\bibliography{ref} 
\end{document}